\documentclass[11pt,final,conference,a4paper]{IEEEtran}
\usepackage[utf8]{inputenc}
\usepackage[T1]{fontenc}
\usepackage[noadjust]{cite}
\usepackage{url}
\usepackage[dvips]{graphicx}

\begin{document}

\IEEEoverridecommandlockouts

\title{Research Directions in Cyber Threat Intelligence}

\author{
	\authorblockN{Stjepan Groš}
 	\authorblockA{
		Faculty of Electrical and Computing Engineering \\
		University of Zagreb\\
		Unska bb, 10000 Zagreb, Croatia \\
 		E-Mail: stjepan.gros@fer.hr}}

\maketitle

\begin{abstract}
Cyber threat intelligence is a relatively new field that has grown
from two distinct fields, cyber security and intelligence. As such,
it draws knowledge from and mixes the two fields. Yet, looking into
current scientific research on cyber threat intelligence research,
it is relatively scarce, which opens up a lot of opportunities.
In this paper we define what cyber threat intelligence is, briefly
review some aspects for cyber threat intelligence. Then, we analyze
existing research fields that are much older that cyber threat
intelligence but related to it. This opens up an opportunity to
draw knowledge and methods from those older field, and in that way
advance cyber threat intelligence much faster than it would by
following its own path. With such an approach we effectively
give a research directions for CTI.
\end{abstract}


\section{Introduction}
\label{sec:intro}

For a long time in human history every state that wanted to survive
had to know \textit{who} its enemies are, \textit{what} are their
capabilities, \textit{when} they intend to do something, \textit{where}
they will do it, \textit{why} they want to do it, and \textit{how}
they intend to do it. These are so called 5W's \cite{burger2014}.
The enemies, of course, could be the other nation states, different
groups of adversaries, but also country's own citizens. For a long
time these kinds of activities were exclusive domain of activity of
nation states and was done by secret services of some kind.


But in the late 20th centry, and especially in 21st century, with
the proliferation of Internet and ever bigger reliance on information
technology (IT) companies found themselves basically in the same
situation nation states find themselves for centuries. 
So, companies started to try to obtain data about their threats,
and basically, started to perform tasks performed by secret services
in order to obtain data about its threats, i.e. 5W's. Yet, companies
are not authorized by law to perform full range of collection activities
performed by secret service, nor they have resources to do so.
Furthermore, cyber domain has its specifics with respect to
domains secret services operate in. So, the two domains were
fused thus creating a new intelligence branch, cyber threat intelligence
(CTI).

The goal of this paper is to review several of the research fields
that the authors identified to have some commonalities with the cyber
threat intelligence, but in the same time are much older than CTI,
with respect to the use, experience and the body of research.
It is the idea that the knowledge
accumulated in those other research fields can be applied to CTI in
some way. In that way we hope to give some guidelines for advancing
cyber threat intelligence much more faster by reusing ideas, methodologies,
experiences and other knowledge elements from the other, older, research
fields. In that way we aim to fulfill the main objective of this paper,
\textit{to give research directions in CTI by connecting it to much
more mature related fields}.

The paper is structured as follows. In the Section \ref{sec:CTI}
we define the necessary terms, the chief among which is cyber threat
intelligence itself. We also give a quick overview of some basic
concepts in the cyber threat intelligence. Then, in Section
\ref{sec:CTIResearchAndRelatedWork} we briefly review research
related specifically to CTI and we also review a related work. 
The Section \ref{sec:TheThreeIntelligences} is about intelligence
already done for some time in today's organizations, i.e.
business intelligence and competitive intelligence. Then, in
Section \ref{sec:RelatedResearchFields} we review some of related
fields of research that fully or in part could be applied in
order to improve CTI processes and its output. Finally, in
Section \ref{sec:ConclusionsAndFutureWork} we give conclusions.

This paper builds on presentation given on the Cybersecurity
sympozium organized by Croatian Ministry of Defense in summer
of 2016 on the island of Mali Lošinj.

\section{Cyber Threat Intelligence}
\label{sec:CTI}

In this section we'll define what CTI is and also describe few
important aspects of the CTI.

\subsection{What is CTI?}

To know what CTI is, we have to define it, and we can start with
the definitions and descriptions of its constitutive parts contained
in the name CTI, i.e. the terms \textit{cyber}, \textit{threat} and
\textit{intelligence}. Then we can try to answer the question posed
in the title of this subsection, i.e. \textit{What is a CTI?}.

First, the word \textit{cyber} denotes domain or environment in which
everything happens. It's usually not used alone, i.e. it is
used as an adjective. In cases when it's used alone it is possible
from the context to infer what it refers to. There are many possible
definitions of the word \textit{cyber} all of which in basically say
the same, cyber is adjective that denotes a special environment.
Here is one such definition \cite{finland2013}: 

	\begin{quote}
	\begin{itshape}
	Its inference usually relates to electronic information
	(data) processing, information technology, electronic
	communications (data transfer) or information and computer
	systems. Only the complete term of the compound word
	(modifier+head) itself can be considered to possess actual
	meaning.
	\end{itshape}
	\end{quote}

The definition lists everything that might be considered "cyber"!
So, for example, events within a computer system is a cyber world,
data while being transferred from source to destination are also
in cyber, etc.

Next, the word \textit{threat} is an umbrella term that includes
anything that can make some harm \textit{to something}. Obviously,
what is considered to be threat depends on that something, i.e.
who you ask! So, in case of a bank, there is one set of threats,
in case of nation state, there's another set of threats, etc.
In conclusion, a very general class of different things can be
regarded as a \textit{threat}, for example earthquake, or
negligent person, or an attacker, etc. Consequently, there are
lot of definitions of the term \textit{threat}, e.g. a
simple one \cite{kissel2013}:

	\begin{quote}\textit{
	The potential source of an adverse event.
	}\end{quote}


Finally, there is the key term \textit{intelligence}. One would think
that for something done for so long there is agreed upon definition,
but the truth is just the opposite! There are multiple definitions,
depending whether you look from the consumer or producer side of
the intelligence \cite{fbi2017}, and if you look traditional
intelligence applied only for the purpose of the nation state,
or the one applied by other entities like different companies.

One modern definition we think is very good and includes not only
the intelligence done by governments and militaries, but also by
companies, is given by Breakspear \cite{breakspear2013}:

	\begin{quote}
	\begin{itshape}
	Intelligence is a corporate capability to forecast change
	in time to do something about it. The capability involves
	foresight and insight, and is intended to identify impending
	change which may be positive, representing opportunity,
	or negative, representing threat.
	\end{itshape}
	\end{quote}

This definition places an emphasis on process, rather then on the
outcome of this process. There are, of course, definitions that don't
put emphasis on one signel aspect of intelligence, like aforementioned
FBI's one \cite{fbi2017}:

	\begin{quote}
	\begin{itshape}
	Intelligence is:

	\begin{itemize}
	\item	Intelligence is a \textbf{product} that consists
		of information that has been refined to meet the
		needs of policymakers;
	\item	Intelligence is also a \textbf{process} through
		which that information is identified, collected,
		and analyzed; and,
	\item	Intelligence refers to both the \textbf{individual
		organizations} that shape raw data into a finished
		intelligence product for the benefit of decision
		makers and the \textbf{larger community of these
		organizations.}
	\end{itemize}
	\end{itshape}
	\end{quote}

To finish with definitions we can now infer from them what
specifically CTI is by applying the meaning of words \textit{cyber}
and \textit{threat} to the word \textit{intelligence}:

\begin{itemize}
\item	It is "traditional" intelligence applied in the realm of cyber
	focusing on threats \cite{chismon2015}.
\item	It is a product produced by a process done by an organization,
	all together called the cyber threat intelligence.
\item	It is done by companies but also government organizations,
	and potentially many others that have established process,
	and produce some product.
\end{itemize}

It should be also noted that there is a big difference between
information and intelligence. This is emphasized in the definition
given by Breakspear who explicitly states that CTI [product] involves
foresight and insight!

To finish with definitions, let us just say that in the literature
it is possible to find two alternative names for Cyber threat
intelligence (CTI), \textit{threat intelligence} \cite{chismon2015} and
\textit{cyber intelligence}. Even though not strictly equivalent if
we look through the lenses of definitions, usually they are equivalent
due to the context in which they are used, i.e. the context provides
for "missing" word, either cyber or threat. There is also term \textit{security
intelligence}, but it is a lot more broader that CTI for a simple
reason that the word \textit{security} means a lot more than
\textit{information security} and \textit{cyber security}.

\subsection{CTI Process}
\label{sec:CTIProcess}

CTI, like intelligence in general, follows a certain process which, as
a result, produces output. There are several models that can be found
in the literature, for example the most basic steps taken from the
\textit{traditional} intelligence are \cite{frini2011}:

\begin{enumerate}
\item	Planing and Direction
\item	Collection
\item	Processing and Exploitation
\item	Analysis and Production
\item	Dissemination and Integration
\end{enumerate}

The goal of \textit{planning and direction} is to define overall goals
of the intelligence process, i.e. intelligence consumers define what
they need. Then in the \textit{collection} phase raw data is collected.
For the collection of raw data traditional intelligence uses many
different techniques, like \textit{open-source intelligence} (OSINT),
\textit{human intelligence} (HUMINT), \textit{signals intelligence}
(SIGINT), and \textit{geospatial intelligence} (GEOINT), to name a few.
But not all of those collection disciplines can be used in CTI by
companies or any other entities other than secret services for the reason
they are illegal and/or non-ethical. So, for CTI the only available
collection discipline is OSINT.

\textit{Processing and exploitation} transforms raw data into information,
and then \textit{analysis and production} turns information into intelligence,
the final product delivered to the intelligence user, which is part of
the \textit{dissemination and integration} steps.

Steps given above are augmented with additional steps that cover
the whole process, like feedback step. Additionally, the process
is drawn like unidirectional, but there are backlinks that make
the process much more iterative whith more smaller cycles. There are a
number of works analyzing and searching for the optimal process,
so we'll not go into details in this paper. 

No matter which exact sequence of steps is taken, it is collectively
known as \textit{intelligence cycle} because steps are usually drawn on a
circle with arrows pointing from one to the next step and the last
step pointing to the first one. The intelligence cycle is not
without its issues, and the collected set of critiques from the
literature can be found in \cite{frini2011}.

\subsection{CTI Levels}

Usually, CTI is divided into three levels, i.e. \textit{strategic},
\textit{operational}, and \textit{tactical}. It is important to emphasize
that there is no strict delineation between the three, i.e. there are
overlaps and sometimes it might be hard to identify for some activity to
which level it belongs to.

At the highest level is \textit{strategic cyber threat intelligence}. It
is closely related to the strategy of the organization and as such its
goal is to support it and protect from potential threats. The product
of the CTI at this level is for organization's CEO and the board that
sets overal strategy and executes it. At the next, lower, level is
\textit{operational cyber threat intelligence}. The goal of CTI at this
level is to identify threats, their tactics, techniques and procedures. All
this allows for building better defenses and as such the consumers of
this intelligence are CISOs and CIOs \cite{mattern2014}. Finally, at the lowest level is
\textit{tactical cyber threat intelligence}. Today it is the most frequently
associated with CTI in general, i.e.  when people talk about CTI in many
cases they actually talk about tactical level and its artifacts
\cite{mattern2014}. For example, indicators of compromise (IoC),
different hash values, description of what malware does and similar are
all the products of tactical level cyber threat intelligence.

Note that it is also possible to find four level classification in the
literature, i.e. \textit{strategic}, \textit{operational},
\textit{tactical} and \textit{technical} \cite{chismon2015}. Still,
the majority of text use three levels and so we adopted that approach, too.

\section{CTI Research and Related Work}
\label{sec:CTIResearchAndRelatedWork}

Research in CTI is currently focused on tactial level, i.e. analysis
of malware, creating indicators of compromise, and sharing them.
Sharing is done using the STIX of OpenIOC formats for describing the
intelligence, and in the case of STIX, TAXII protocol for the exchange
of the information. Much less information is available on operational
level of cyber threat intelligence. And even less so about strategic
intelligence. That doesn't mean there is no reports covering all the
aspects of cyber threat intelligence. Different threat reports
published by companies like FireEye, Kaspersky, Sophos, IBM, etc. contain
intelligence on strategic, operational and tactical levels, but the
emphasis is on tactical and less on operational levels. Furthermore,
the information on how the intelligence was collected, processed, etc.
is unavailable and probably mainly done automatically on techical
level, and manually on all the levels.

As for the similar work to the research reported in this paper, the
authors aren't aware of any. So, this seems to be a promising direction
with a lot of oportunity for further, and much deeper, research not
yet done by others.

\section{The Three Intelligences}
\label{sec:TheThreeIntelligences}

The universe of any company consists of the following stakeholders:
(\textit{i}) customers, (\textit{ii}) competitors, (\textit{iii})
partners, and (\textit{iv}) attackers. Each one of them interacts with
the company on every level (tactical, operational, strategic),
and thus influences the state the company
will be in short, medium and long term, respectively. For that reason companies
developed means of tracking stakeholders and, in a way, predicting
what will happen. Competitors are tracked using competitive intelligence
(CI), customers, the company itself and other data using business intelligence (BI), and threats/attackers
using cyber threat intelligence, (CTI). Partners, if necessary, can be
tracked using CI or BI, depending on the exact nature of partnership.

Of the three, CTI is a relative new comer to the game, i.e. BI and CI
are practiced for a lot longer time. This means that the two accumulated
a lot more knowledge in due time, and due to certain similarities, very
likely this knowledge can be applied for CTI as well, advancing it more
rapidly than by going through all development/research steps BI and CI
went over the years. In the following sections we'll talk in a bit more
details about CI and BI.

\subsection{Competitive Intelligence}
\label{sec:CompetitiveIntelligence}

Companies are performing competitive intelligence in some form for a
long time now, at least 30 years now. The goal of \textit{competitive
intelligence} (CI) is to collect information about competition, analyze
it and present managers with actionable data \cite{bose2008}. The CI is
especially important today in a highly competitive market where decision
makers have to have actionable data in order to properly guide companies
and where mistakes and wrong predictions can have disastrous effects.
For that reason many companies have established competitive intelligence
processes. Furthermore, the fact that this activity exists for a relatively
long time, as well as its importance, means that substantial body of
knowledge has been accumulated with time. 

\subsection{Business Intelligence}
\label{sec:BusinessIntelligence}

Business Intelligence is relatively old activity going back as far as
to 1950ties \cite{luhn1958} even though the true renaissance the field
achieved in 90ties. The goal of BI is to analyze information present
withing the company and to extract intelligence (knowledge) that can be
used to support decision making by management. The main difference 
between BI and CI is that former uses internal information, while
latter uses external information. For a long time the primary data
sources for BI were structured databases, but lately unstructured
text is also used due to having vast amounts of it available in comparison to
structured data. Furthermore, the advancement of techniques to analyze
unstructured data like natural language processing and machine intelligence
also helped in that respect.

\section{Related Research Fields}
\label{sec:RelatedResearchFields}


The business intelligence and competitive intelligence described in the
previous section use many methods to analyze and process data. Some of
those are described here along with some additional tools that might be
used in the CTI. The list is in alphabetical order.

\subsection{Adversarial Reasoning}
\label{sec:AdversarialReasoning}

\textit{Adversarial reasoning} is subfield of Decision making and its
goal is to model and predict how adversarials will behave
\cite{pelta2009}. This in turn allows one to properly respond to
anticipated behaviors.  The field of adversary reasoning is relatively
old, having first papers appeared in 1992 \cite{pelta2009}, so in
the mean time it is obviously accumulated a lot of knowledge.
The main tool of adversarial reasoning was game theory, but that isn't
the case any more. The connection with the CTI is obvious, namely, CTI
analyst when analyzing threats has to anticipate threat's behavior
in medium (operational) and long (strategic) time frame.

\subsection{Cognitive Computing}
\label{sec:CognitiveComputing}

Cognitive computing is a relatively new field but it is a continuation
of an old idea of Artificial Intelligence \cite{langley2012}. Namely, 
the original idea of AI was to mimic human mind in it capacity to
comprehend and understand the world. This turned out to be very hard
and with the time AI fragmented into multiple subfields each with its
own narrow goals. Yet, all of those fields carry a mark of AI and thus
AI itself is a very overloaded expression. To avoid all those conotations
by using the name AI, \textit{Cognitive Computing} term was coined.
The idea is to use multiple methods in parallel (sequentially) and achieve
more complex behaviors than it is possible using only a single method, like
deep learning, etc. Since the field is relatively new it is still
confusing, and that is the most noticeable by comparing scientific
literature like \cite{langley2012} and commercial like \cite{high2012a}.
Commercial ones put an emphasis on processing natural language, while
scientific ones take a more broader approach. But, either way, the
biggest success of cognitive computing to date is IBM Watson \cite{high2012a}
which in 2011 managed to beat two of the best players in Jeopardy quiz show \cite{jeopardywatson}.
This is a significant success taking into account how hard is for computers
to understand ambiguous human language, and to do so quickly. After that
success IBM decided to commercialize IBM Watson and offer it as a cloud
service to the customers.

CTI would obviously benefit from cognitive computing in general, and
IBM Watson like systems in particular, because they would allow
harvesting \textit{and understanding} a large amounts of data, something
that is today very expensive and very hard to achieve, if possible at
all.

%

\subsection{Information Extraction}
\label{sec:InformationExtraction}

\textit{Information extraction} is a research discipline that tries to
solve the problem of automated extraction of information from unstructured
texts, like documents, Web pages, forums, etc. It grew up from natural
language processing, and is developed for over 25 years now \cite{sarawagi2008}.

\subsection{Data and Information Fusion}
\label{sec:DataAndInformationFusion}

The field of \textit{data and information fusion} deals with the fusion
of information coming from different sources and on a different abstraction
levels in such a way to produce new, more complete and better data,
information, or decisions, at a level of a human understanding then
it would be possible by using the sources separately \cite{foo2013,corona2009}.

Several points are worth emphasizing that connect DIF to CTI. First,
there are different models proposed for information fusion which 
broadly can be grouped by input and output, i.e. input can be either
data, features or decisions (from lowest to highest level), while
the output can be anything of the same or higher level. For example,
if the inputs are features, then the output can be features or
decisions. The next point worth emphasizing is the fact that
intelligence cycle described in the Section \ref{sec:CTIProcess}
is only one model proposed for the purpose of information fusion
\cite{foo2013}. Others are OODA loop, Waterfall model, the
Omnibus model, the Ensley model, and many others. So, the knowledge
from DIF can be applied to the specific parts of the CTI cycle, or
to the whole cycle in general. Also of interest is that the emphasis
on research in the field of DIF is now on high-level fusion. Since
the emphasis of current CTI research and practice is concentrated on
technical/tactical details it might be inferred that CTI is currently
in early stages of development when compared to DIF.

The use of DIF in computer and network security has already been
suggested. E.g., Corona et. al \cite{corona2009} review the 
use of DIF for the purpose of improving intrusion detection.

\subsection{Situational Awareness and Cyber Situational Awareness}
\label{sec:SituationalAwareness}

The \textit{situational awareness} (SAW) is \textit{the perception of
the elements in the environment within a volume of time and space,
the comprehension of their meaning, and the projection of their
status in the near future} \cite{foo2013}. The \textit{Cyber Situational
Awareness} (CSAW) is concerned with situational awareness in the cyber
space. SAW is dependent on information and DIF so the two are often
interrelated in the literature.

Obviously the goal of CTI is to make decision makers aware of threats
in cyber space, which in other words, is very similar to cyber
situational awareness, though a bit more narrow.

\subsection{Knowledge Management}
\label{sec:KnowledgeManagement}

\textit{Knowledge management} (KM) is a discipline that
provides strategy, process, and technology to share and
leverage information and expertise that will increase our
level of understanding to more effectively solve problems
and make decisions \cite{satyadas2001}.

KM is about managing a \textit{knowledge}, which is defined
in variety of ways, one of which is that it's
\textit{actionable information} \cite{satyadas2001}. This
is in line with what CTI produces, i.e. CTI from data
extracts information and from information produces knowledge.
Thus, it is interesting to see what knowledge management is,
and what can be (re)used to improve CTI and its processes.

There are a number of reasons businesses use KM \cite{satyadas2001}.
Almost all of which have application for the purpose of CTI, too.
Here are just a few with the justification on why they should
be used for CTI:

\begin{itemize}
\item	\textit{measure and track value and flow of information} --
	this is a very important issue because when something is
	better measured, the easier it is to manage it. There is
	no engineering and science without measurement. This holds
	for everything, and thus for CTI, too.

\item	\textit{the need to build highly effective virtual teams
	and reduce cycle times, as well as better customer
	relationships} -- again, something very important for
	today's organizations that helps the organization to
	manage and improve its processes, so applicable to CTI,
	too.

\item	\textit{ the need for a better way to share information
	and knowledge across organizational boundaries and the
	ability to rapidly respond to queries or crisis} --
	information sharing is a hot topic in information security
	in general, and in CTI in particular. As can be seen,
	this is a topic with which KM deals in a methodological
	way, so this makes KM very applicable to CTI and Information
	security in general!
\item	\textit{the need to retain the knowledge of experts
	who are retiring} -- This is the problem that has
	every organization with its workforce. Not only when experts
	are retiring, but also if they leave a company for
	other job. Keeping their knowledge is very valuable
	for the continuity of processes done within the
	organization.
\end{itemize}

There are other business reasons for using KM in organizations
that could be also applied for CTI. But this much is enough to
argue that KM is a very interesting field of research with a
lot of potential to improve CTI and its processes.

%
%
%
%
%
%
%
%


%
%
%
%
%
%
%

%

\section{Conclusions and Future Work}
\label{sec:ConclusionsAndFutureWork}

In this paper we defined CTI and gave a quick overview of CTI methods
and processes. We then reviewed related research fields, starting with
business and competitive intelligence and continuing with other
related fields. We argued that each research field has something in
common with the CTI thus, meaning that CTI can be advanced much
quicker by reusing some knowledge from those related fields. If we
take into account that each of the related fields is much older (in
time frames of information technology) than CTI, then the potential
is really big. In that way we showed the potential research directions
for CTI.

The work in this paper is just a start since there are much more research
topics that has to be done. First, the relation between research
fields presented in this paper and CTI can be correlated much deeper,
and also there are other fields with a potential to be (re)used
in CTI, like \textit{knowledge engineering}, \textit{Ontologies},
\textit{Uncertainty Analysis/Management}, \textit{Decision Theory},
\textit{Critical Thinking}, \textit{Uncertainty Theory \& Reasoning},
\textit{Big Data}, and \textit{Cognition}.

\bibliographystyle{IEEEtran}
\bibliography{bibliography}

\begin{thebibliography}{10}
\providecommand{\url}[1]{#1}
\csname url@samestyle\endcsname
\providecommand{\newblock}{\relax}
\providecommand{\bibinfo}[2]{#2}
\providecommand{\BIBentrySTDinterwordspacing}{\spaceskip=0pt\relax}
\providecommand{\BIBentryALTinterwordstretchfactor}{4}
\providecommand{\BIBentryALTinterwordspacing}{\spaceskip=\fontdimen2\font plus
\BIBentryALTinterwordstretchfactor\fontdimen3\font minus
  \fontdimen4\font\relax}
\providecommand{\BIBforeignlanguage}[2]{{%
\expandafter\ifx\csname l@#1\endcsname\relax
\typeout{** WARNING: IEEEtran.bst: No hyphenation pattern has been}%
\typeout{** loaded for the language `#1'. Using the pattern for}%
\typeout{** the default language instead.}%
\else
\language=\csname l@#1\endcsname
\fi
#2}}
\providecommand{\BIBdecl}{\relax}
\BIBdecl

\bibitem{burger2014}
E.~W. Burger, M.~D. Goodman, P.~Kampanakis, and K.~A. Zhu, ``Taxonomy model for
  cyber threat intelligence information exchange technologies,'' in
  \emph{Proceedings of the 2014 ACM Workshop on Information Sharing \&
  Collaborative Security}.\hskip 1em plus 0.5em minus 0.4em\relax ACM, 2014,
  pp. 51--60.

\bibitem{finland2013}
{Secretariat of the Security and Defence Committee}, ``{Finland's Cyber
  security Strategy},'' Finland, 2013.

\bibitem{kissel2013}
R.~Kissel, ``Glossary of key information security terms,'' \emph{NIST
  Interagency Reports NIST IR}, vol. 7298, no.~3, 2013.

\bibitem{fbi2017}
``{FBI Intelligence Branch},''
  \url{https://www.fbi.gov/about/leadership-and-structure/intelligence-branch},
  accessed: 2017-02-19.

\bibitem{breakspear2013}
A.~Breakspear, ``A new definition of intelligence,'' \emph{Intelligence and
  National Security}, vol.~28, no.~5, pp. 678--693, 2013.

\bibitem{chismon2015}
D.~Chismon and M.~Ruks, ``Threat intelligence: Collecting, analysing,
  evaluating,'' 2015, {MWR Infosecurity, UK Cert, United Kingdom}.

\bibitem{frini2011}
A.~Frini and A.-C. Boury-Brisset, ``An intelligence process model based on a
  collaborative approach,'' DTIC Document, Tech. Rep., 2011.

\bibitem{mattern2014}
\BIBentryALTinterwordspacing
T.~Mattern, J.~Felker, R.~Borum, and G.~Bamford, ``Operational levels of cyber
  intelligence,'' \emph{International Journal of Intelligence and
  CounterIntelligence}, vol.~27, no.~4, pp. 702--719, 2014. [Online].
  Available: \url{http://dx.doi.org/10.1080/08850607.2014.924811}
\BIBentrySTDinterwordspacing

\bibitem{bose2008}
R.~Bose, ``Competitive intelligence process and tools for intelligence
  analysis,'' \emph{Industrial Management \& Data Systems}, vol. 108, no.~4,
  pp. 510--528, 2008.

\bibitem{luhn1958}
H.~P. Luhn, ``A business intelligence system,'' \emph{IBM Journal of Research
  and Development}, vol.~2, no.~4, pp. 314--319, 1958.

\bibitem{pelta2009}
D.~A. Pelta and R.~R. Yager, ``Dynamic vs. static decision strategies in
  adversarial reasoning.'' in \emph{IFSA/EUSFLAT Conf.}\hskip 1em plus 0.5em
  minus 0.4em\relax Citeseer, 2009, pp. 472--477.

\bibitem{langley2012}
P.~Langley, ``The cognitive systems paradigm,'' \emph{Advances in Cognitive
  Systems}, vol.~1, pp. 3--13, 2012.

\bibitem{high2012a}
R.~High, ``The era of cognitive systems: An inside look at {IBM Watson} and how
  it works,'' \emph{IBM Corporation, Redpapers}, 2012.

\bibitem{jeopardywatson}
\BIBentryALTinterwordspacing
{IBM Corporation}. (2011) {Dave Ferrucci at Computer History Museum: How it all
  began and what's next}. [Online]. Available:
  \url{https://www.ibm.com/blogs/research/2011/12/dave-ferrucci-at-computer-history-museum-how-it-all-began-and-whats-next/}
\BIBentrySTDinterwordspacing

\bibitem{sarawagi2008}
S.~Sarawagi \emph{et~al.}, ``Information extraction,'' \emph{Foundations and
  Trends{\textregistered} in Databases}, vol.~1, no.~3, pp. 261--377, 2008.

\bibitem{foo2013}
P.~H. Foo and G.~W. Ng, ``High-level information fusion: An overview.''
  \emph{Journal of Advances in Information Fusion}, vol.~8, no.~1, pp. 33--72,
  2013.

\bibitem{corona2009}
I.~Corona, G.~Giacinto, C.~Mazzariello, F.~Roli, and C.~Sansone, ``Information
  fusion for computer security: State of the art and open issues,''
  \emph{Information Fusion}, vol.~10, no.~4, pp. 274--284, 2009.

\bibitem{satyadas2001}
A.~Satyadas, U.~Harigopal, and N.~P. Cassaigne, ``Knowledge management
  tutorial: an editorial overview,'' \emph{IEEE Transactions on Systems, Man,
  and Cybernetics, Part C (Applications and Reviews)}, vol.~31, no.~4, pp.
  429--437, 2001.

\end{thebibliography}

\end{document}